\documentclass[prl,reprint,amsmath,amssymb,floatfix,citeautoscript]{revtex4-1}
\usepackage{cancel}
\usepackage{amsmath}
\usepackage{mathtools}
\usepackage{graphicx}
\usepackage{amssymb}
\usepackage{amsthm}
\usepackage{bm}
\usepackage{dcolumn}

\usepackage{ragged2e}
\usepackage[labelsep=period,format=plain,justification=justified,font=footnotesize]{caption}
\DeclareCaptionJustification{myjust}{\justifying}
\usepackage{txfonts}

\usepackage{color}
\usepackage[usenames,dvipsnames]{xcolor}
\definecolor{myblue}{rgb}{0,0,1}
\usepackage[breaklinks=true,colorlinks=true,linkcolor=myblue,urlcolor=myblue,citecolor=myblue]{hyperref}

\let\vr\undefined
\newcommand{\vr}{{\bm{r}}}

\newcommand{\vrho}{{\bm{\rho}}}
\newcommand{\vk}{{\bm{k}}}

\newcommand{\vq}{{\bm{q}}}

\newcommand{\vG}{{\bm{G}}}

\newcommand{\vx}{{\bm{x}}}

\newcommand{\eps}{\varepsilon}

\begin{document}

\title{Environmentally-Sensitive Theory of Electronic and Optical Transitions\\
in Atomically-Thin Semiconductors}

\author{Yeongsu Cho}
\affiliation{Department of Chemistry and James Franck Institute,
University of Chicago, Chicago, Illinois 60637, USA}

\author{Timothy C. Berkelbach}
\email{berkelbach@uchicago.edu}
\affiliation{Department of Chemistry and James Franck Institute,
University of Chicago, Chicago, Illinois 60637, USA}

\begin{abstract}
We present an electrostatic theory of band gap renormalization in
atomically-thin semiconductors that captures the strong sensitivity to the
surrounding dielectric environment.  In particular, our theory aims to correct
known band gaps, such as that of the three-dimensional bulk crystal.  Combining
our quasiparticle band gaps with an effective mass theory of excitons yields
environmentally-sensitive optical gaps as would be observed in absorption or
photoluminescence.  For an isolated monolayer of MoS$_2$, the presented theory
is in good agreement with \textit{ab initio} results based on the $GW$
approximation and the Bethe-Salpeter equation.  We find that changes in the
electronic band gap are almost exactly offset by changes in the exciton binding
energy, such that the energy of the first optical transition is nearly
independent of the electrostatic environment, rationalizing experimental
observations.  
\end{abstract}

\maketitle

\textit{Introduction.}
Atomically-thin materials exhibit remarkable electronic properties due to their
quasi-two-dimensional
nature~\cite{Mak2008,CastroNeto2009,Splendiani2010,Mak2010}.  However, their
size also makes them extremely sensitive to their local environment.  A
complete theoretical picture must simultaneously treat the two-dimensional
nature of carriers and the dielectric character of the surroundings. This
latter property is the primary distinction between atomically-thin materials
(such as the transition metal dichalcogenides) and heterostructured
semiconductor quantum wells (such as GaAs in AlGaAs).

To date, many theoretical studies of atomically-thin materials have focused on
the excitonic properties, including the large exciton binding
energy~\cite{Berkelbach2013,Berghauser2014,Olsen2016}, the unique excitonic
Rydberg series~\cite{Chernikov2014,Qiu2013}, the nature of selection
rules~\cite{Berkelbach2015,Gong2017,Cao2017}, and Berry phase modifications of
the exciton spectrum~\cite{Srivastava2015,Zhou2015}.  Surprisingly, the
quasiparticle band gap has received significantly less attention, especially
from simplified microscopic theories, perhaps because it is challenging to
measure experimentally.  In fact, simple theories of the exciton binding energy
are often times used in conjunction with the experimentally measured optical
gap in order to estimate the quasiparticle band
gap~\cite{Chernikov2014,Raja2017}.

The $GW$ approximation represents the current method-of-choice for the accurate
calculation of band structures and band gaps~\cite{Hedin1965,Hybertsen1985}.
However, the quasi-two-dimensional nature of the atomically-thin materials
makes these calculations very challenging to
converge~\cite{Komsa2012,Huser2013,Qiu2016}.  In this work, we provide a simple
electrostatic theory of band gap renormalization due to electrostatic proximity
effects.  Through combination with an effective mass theory of the exciton
binding energy, we find that the optical gap -- i.e.~the sum of the band gap
and the (negative) exciton binding energy -- is extremely insensitive to the
dielectric environment.  To the best of our knowledge, this represents the
first quasi-analytical demonstration of this remarkable effect.

\label{sec:theory}

\begin{figure}[b]
\centering
\includegraphics[scale=1.0]{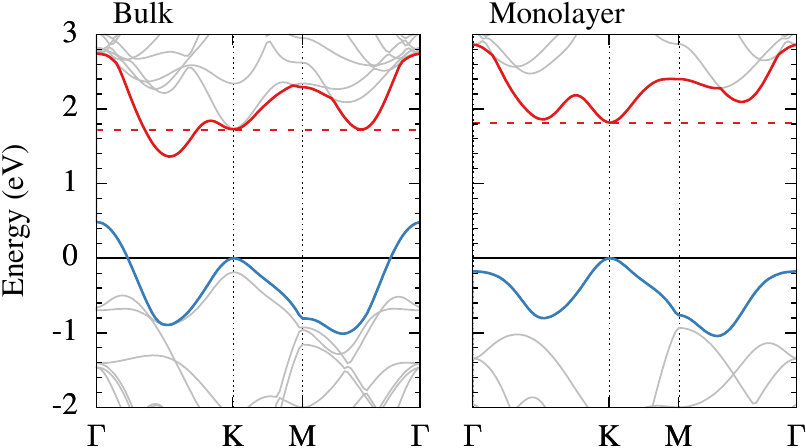}
\caption{
Band structure of bulk and monolayer MoS$_2$ calculated with density functional
theory.  The direct band gap (at the K-point) is 0.09~eV larger for the
monolayer than for the bulk, due to the carrier confinement effect.
}
\label{fig:bands}
\end{figure}

The band gap of nanoscale materials differs from that of the bulk parent
material because of two separate effects: carrier confinement and dielectric
contrast.  In the first case, the geometric confinement of carriers leads to an
increased kinetic energy and a concomitantly larger band gap.  However, in
layered materials (such as the TMDCs), the two-dimensional confinement is
already largely reflected in the bulk band gap, as evidenced by the small
bandwidth in the perpendicular (stacking) direction.  Therefore, in the
following, we employ this idealized scenario of carriers confined to two
dimensions, even when describing the bulk material.
In particular, this approximation is invoked to describe low-energy carriers at
the K-points of the Brillouin zone; here, the wavefunction character is
primarily that of transition-metal $d$-orbitals, which are confined to the
center of the TMDC layer, precluding strong interlayer hybridization.  In
Fig.~\ref{fig:bands}, we show the bandstructure of bulk and monolayer MoS$_2$
calculated using density functional theory~\footnote{Density functional theory
calculations were performed with the Quantum ESPRESSO software
package~\cite{Giannozzi2009}, with the PBE exchange-correlation
functional~\cite{Perdew1996}, norm-conserving pseudopotentials, and a $12\times
12\times 1$ ($12\times 12\times 3$) sampling of the Brillouin zone for the
monolayer (bulk).}. 
The monolayer band gap at the K-point is only 0.09~eV larger than that of the
bulk, indicating that any band gap renormalization due to carrier confinement is
already (largely) accounted for in the bulk band gap; we henceforth neglect this
small shift so as to focus on alternative effects while treating the monolayer
and bulk on equal footing.  We emphasize that this geometric carrier
confinement is a one-electron (kinetic energy) effect that is well-described by
density functional theory -- unlike dielectric screening effects.

\begin{figure}[t]
\centering
\includegraphics[scale=1.0]{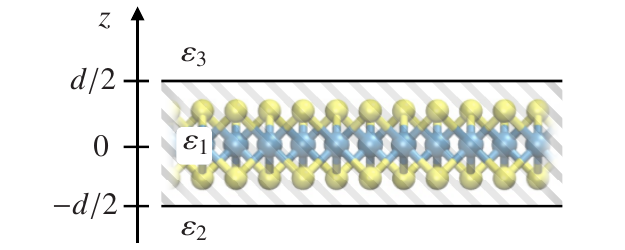}
\caption{
Idealized dielectric slab geometry used to model the electrostatics of
atomically-thin semiconductors.
}
\label{fig:scheme}
\end{figure}

As mentioned above, a second source of band gap renormalization in
nanomaterials is the dielectric contrast effect.  Physically, we recall that
the quasiparticle conduction and valence bands measure the electron affinities
and ionization potentials, respectively.  The excess charge created in these
processes polarizes the material and its environment such that the potential
energy of the charge depends on the local dielectric geometry.  We model
atomically-thin semiconductors as a slab of dielectric constant $\eps_1$ and
width $d$, surrounded by environmental dielectric constants $\eps_2$ below and
$\eps_3$ above, as shown in Fig.~\ref{fig:scheme}.  Consistent with the
arguments presented above, the carriers will be assumed to occupy the center of
the slab, at $z=0$.

We now proceed to calculate the band structure corrections due to such a
heterogeneous dielectric environment.  We assume that a \textit{reference}
many-body band gap is known, which could come from experiment or calculation.
In particular, we will primarily consider band structure corrections to the
three-dimensional bulk material.  Corrections will be calculated in two ways:
(1) classically, using electrostatic continuum theory; and (2) quantum
mechanically, using the static Coulomb-hole plus screened exchange (COHSEX)
approximation to the quantum mechanical $GW$ self-energy.  When correcting a
reference band structure, we require the \textit{difference} in the screened
Coulomb interaction, $\delta W(\vr,\vr^\prime) \equiv W(\vr,\vr^\prime) -
W^{\mathrm{ref}}(\vr,\vr^\prime)$, where $W$ is the total screened Coulomb
interaction.  We calculate the respective screened interactions through their
electrostatic counterparts associated with the slab dielectric geometry shown
in Fig.~\ref{fig:scheme}.  While this is a classical approximation, which
neglects local field effects, it avoids the high cost of an ab initio
calculation of the screened Coulomb interaction.

In recent years, effective mass theories of atomically-thin materials have made
frequent use of the model potential energy derived by Rytova~\cite{Rytova1967}
and Keldysh~\cite{Keldysh1979} (RK),
\begin{equation}
W^{\mathrm{RK}}(\rho) = \frac{\pi e^2}{(\eps_2 + \eps_3) \rho_0}
    \left[ H_0\left(\frac{\rho}{\rho_0}\right) 
        - Y_0\left(\frac{\rho}{\rho_0}\right)\right]
\end{equation}
where $H_0$ and $Y_0$ are the Struve function and the Bessel function of the
second kind and $\rho$ is the two-dimensional in-plane separation.  The
screening length is given by $\rho_0 = \eps_1 d/(\eps_2+\eps_3)$ and can be
related to a two-dimensional sheet
polarizability~\cite{Cudazzo2011,Berkelbach2013}.  For the purposes of the
present manuscript, the RK potential suffers from two deficiencies.  First, it
applies only in the limit of extreme dielectric mismatch between the slab and
its surroundings; while this approximation is good for isolated (suspended)
monolayers, it breaks down in more general dielectric environments.  Second,
the RK potential has an unphysical logarithmic divergence at $\rho = 0$, which
precludes its use in simple electrostatic theories of band gap renormalization.
Instead, we employ the exact solution of the finite-thickness electrostatic
problem shown in Fig.~\ref{fig:scheme}.  We emphasize that the logarithmic
behavior of the RK potential is correct over some intermediate length scale and
only incorrect for $\rho \lesssim d$.

The potential energy of two charges in a slab with locations $z_1, z_2$, and
in-plane separation $\rho$ can be calculated via image charges to give a
screened interaction $W(z_1,z_2,\rho)$~\cite{Kumagai1989}.  In the center of
the slab ($z_1=z_2 =0$), we find
\begin{equation}
\begin{split}
\label{eq:W}
W(\rho) &= \frac{e^2}{\eps_1 \rho}
    + 2\sum_{n=1}^{\infty} \frac{e^2 L_{12}^{n} L_{13}^{n}}
        {\eps_1 \left\{\rho^2 
            + (2nd)^2 \right\}^{1/2}} \\
    &\hspace{1em}
    +(L_{12}+L_{13})\sum_{n=0}^{\infty} \frac{e^2 L_{12}^n L_{13}^n}
        {\eps_1 \left\{\rho^2 
            + [(2n+1)d]^2 \right\}^{1/2}}
\end{split}
\end{equation}
where $L_{1n} = (\eps_1 - \eps_n)/(\eps_1 + \eps_n)$.  Unlike the RK potential,
this continuum electrostatic potential is correct in the uniform case $\eps_1 =
\eps_2 = \eps_3$ and has the proper divergence as $\rho \rightarrow 0$.

\textit{Electrostatic solution.}
In the simplest electrostatic (Born) approximation, the conduction and valence
band corrections in the center of the slab are given by the self-interaction
energy~\cite{Brus1983,Kumagai1989}
\begin{equation}
\label{eq:sigma}
\delta \Sigma_{\mathrm{c/v}} = \pm \frac{1}{2} 
    \lim_{\rho \rightarrow 0} \delta W(\rho),
\end{equation}
which is non-divergent due to the use of an interaction difference, $\delta W$,
as long as the slab dielectric $\eps_1$ is identical in both $W$ and
$W^{\mathrm{ref}}$.  
When the reference potential energy is that of a uniform, bulk dielectric,
i.e.~$W^{\mathrm{ref}}(\vr,\vr^\prime) = e^2/(\eps_1 |\vr-\vr^\prime|)$, then
the electrostatic corrections using Eqs.~(\ref{eq:W}) and (\ref{eq:sigma}) can
be summed analytically to give the relatively simple expression
\begin{equation}
\begin{split}
\label{eq:electro}
\delta \Sigma_{\mathrm{c/v}} &= \pm \frac{e^2}{2\eps_1 d}
    \Bigg\{\frac{2(L_{12}+L_{13})}{\sqrt{L_{12}L_{13}}}
        \tanh^{-1}\left(\sqrt{L_{12}L_{13}}\right) \\
    &\hspace{5em} - \log(1-L_{12}L_{13}) \Bigg\}.
\end{split}
\end{equation}

\textit{Tight-binding COHSEX.}
First-principles band structure calculations typically employ the $GW$
approximation to the self-energy.  In the static screening limit, this
approximation yields two contributions to the self-energy: a Coulomb-hole (COH)
term and a screened exchange (SEX) term~\cite{Hedin1965}.  By assuming that an
initial, many-body \textit{reference} band structure is known, we can calculate
corrections in alternative electrostatic environments as diagonal elements of
the self-energy operator, which leads to
\begin{subequations}
\begin{align}
\delta\Sigma^\mathrm{COH}_p(\vk) &= \frac{1}{2} \lim_{\rho \rightarrow 0} \delta W(\rho), \\
\begin{split}
\delta\Sigma^\mathrm{SEX}_p(\vk) &= - \frac{1}{N_k}\int d^2\vx_1 \int d^2 \vx_2 \phi^*_{p,\vk}(\vx_1) 
    \rho(\vx_1,\vx_2) \\
    &\hspace{8em} \times \delta W(\rho) \phi_{p,\vk}(\vx_2),
\end{split}
\end{align}
\end{subequations}
where $\vx = (\vrho,\tau)$ is the combined space and spin variable,
$\rho(\vx_1,\vx_2)$ is the reduced density matrix of the mean-field reference,
$N_k$ is the number of $k$-points sampled in the Brillouin zone,
and $p = (\mathrm{c,v})$ indexes the conduction or valence band.
In the simplest approximation, we consider the two-band tight-binding 
Hamiltonian~\cite{Xiao2012}
\begin{equation}
\label{eq:tb}
H(\vk) = \left( \begin{array}{cc}
E_\mathrm{g}/2 & at(k_x+ik_y) \\
at(k_x-ik_y) & -E_\mathrm{g}/2
\end{array} \right)
\end{equation}
with eigenvectors $\langle\vx|p\vk\rangle=\phi_{p\vk}(\vx)$ and eigenvalues 
$E_{\mathrm{c/v}}(\vk) = \pm \frac{1}{2}\sqrt{E_g^2 + (2 a t k)^2}$. 
In this Hamiltonian, $E_{\mathrm{g}}$ is the band gap, $a$ is the lattice
constant, and $t$ is the interatomic transfer integral.
A single (doubly-occupied) valence band leads to the simple density matrix
$\rho(\vx_1,\vx_2) = \sum_{\vq} \phi_{v\vq}(\vx_1)\phi_{v\vq}^*(\vx_2)$.
Further simplifications concerning the locality of the underlying real-space
basis functions leads to the SEX self-energy
\begin{equation}
\label{eq:SEX_tb}
\delta\Sigma^\mathrm{SEX}_p(\vk) = - \frac{1}{N_k}\sum_\vq 
|\langle p\vk|v\vq\rangle|^2 
\sideset{}{^\prime}\sum_{\vG} \delta W(\vG + \vq-\vk),
\end{equation}
where
\begin{equation}
\delta W (\vk) =  \frac{1}{A_{\mathrm{BZ}}}\int d^2\rho\ e^{i\vrho\cdot\vk}\delta W(\rho),
\end{equation}
$A_{\mathrm{BZ}}$ is the area of the Brillouin zone,
and the primed summation in Eq.~(\ref{eq:SEX_tb}) excludes the term with $\vG=0$ when
$\vk=\vq$.
Summarizing, the COH term yields a positive, constant shift to both the conduction and
valence band, which is \textit{exactly} equal to the (positive) correction obtained
in the pure electrostatic theory presented above; the SEX term yields a negative,
$k$-dependent shift with a magnitude that depends on overlap factors between the
valence band and the band being corrected.  To a reasonable approximation (verified
numerically below), the SEX contribution is negligible in the conduction band
(due to vanishing overlaps) but is substantial in the valence band.
Further, if the squared overlap is approximated by unity, 
i.e.~$|\langle v\vk|v\vq\rangle|^2 \approx 1$, then the magnitude of the
SEX correction in the valence band is exactly twice that of the COH term.  As shown
in Ref.~\onlinecite{Neaton2006} for the case of molecules near metal surfaces,
we therefore have simple, approximate COHSEX corrections given by
$\delta\Sigma_\mathrm{c} \approx +P - 0 = +P$ and 
$\delta\Sigma_\mathrm{v} \approx +P - 2P = -P$,
where $P = \frac{1}{2}\lim_{\rho\rightarrow 0} \delta W(\rho)$
is precisely the electrostatically-derived correction.  
In reality, the squared overlap
can be less than one, and the SEX correction to the valence band (and thus the
band gap) will be slightly smaller than that of the continuum electrostatic
theory.

\textit{Effective-mass theory of excitons.}
The optical gap, as measured in linear spectroscopies such as absorption
or photoluminescence, is the sum of the quasiparticle band gap and
the (negative) exciton binding energy.  At a similar level of theory
to that used so far, the exciton states can be calculated
using an effective mass theory,
\begin{equation}
\left[ -\frac{1}{2\mu}\nabla_\vrho^2 - W(\rho) \right] \Psi_n(\vrho)
    = E_n \Psi_n(\vrho),
\end{equation}
where $\rho$ is the electron-hole separation, $\Psi_n$ is the exciton
wavefunction, and $E_n$ is its binding energy.  The material parameters enter
through 
the exciton reduced mass
$\mu = m_{\mathrm{e}}m_{\mathrm{h}}/(m_{\mathrm{e}}+m_{\mathrm{h}})$
and the same screened Coulomb interaction $W$ as used above. Due to the angular
symmetry, the effective mass equation is a simple one-dimensional Schr\"odinger
equation in the radial direction, which may be solved numerically exactly on a
real-space grid to obtain the full Rydberg series of band-edge excitons.  The
exciton wavefunctions and binding energies are sensitive to the local
dielectric environment, where higher dielectric constants result in stronger
screening, more diffuse wavefunctions, and smaller binding energies.

\textit{Results.}
While our theory is appropriate for any atomically-thin semiconductor,
we will apply it to the well-studied case of MoS$_2$, a prototypical
layered transition-metal dichalcogenide.
As is common for quantum-confined materials, we correct the bulk band gap using
a uniform reference Coulomb potential with $\eps_1 = \eps_2 = \eps_3$,
i.e.~$W^{\mathrm{ref}}(\vr,\vr^\prime) = e^2/(\eps_1 |\vr-\vr^\prime|)$~\footnote{In 
reality, the screening of bulk
TMDCs is anisotropic, and one could imagine using a more complicated, but
realistic reference potential; however, using a coarse-grained treatment of the
anisotropy, $W^{\mathrm{ref}}(\vr,\vr^\prime) = e^2/[\eps_z \eps_{xy} \rho^2 +
\eps_{xy}^2(z-z^\prime)^2]^{1/2}$, gives the same result in the order
$z\rightarrow z^\prime$ with $\rho=0$.};
for MoS$_2$, we use $\eps_1=14$.
For the monolayer, we solve the electrostatic problem in Fig.~\ref{fig:scheme} with 
$\eps_1=14$ and $d = 6$~\AA, which roughly corresponds to the perpendicular extent 
of monolayer MoS$_2$; these parameters yield the ideal screening length $\rho_0
= 42$~\AA\, in good agreement with the ab initio value of
41.5~\AA~\cite{Berkelbach2013}.  We take the reference A-series band gap of
bulk MoS$_2$ to be $E_{\mathrm{g}}^{\mathrm{bulk}} = 1.98$~eV~\cite{Evans1965}
and for the tight-binding Hamiltonian in Eq.~(\ref{eq:tb}), we use $at =
3.51$~eV$\cdot$\AA.  

\begin{figure}[b]
\centering
\includegraphics[scale=1.0]{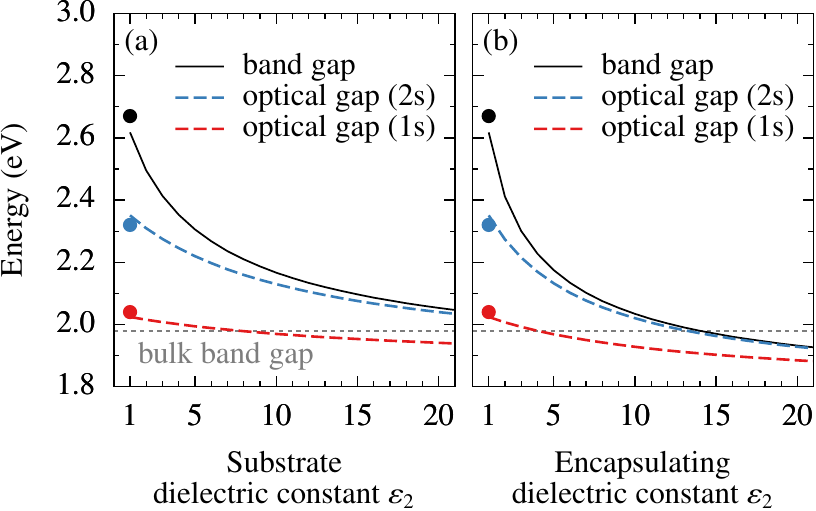}
\caption{
Quasiparticle band gap and optical gap (i.e.~excitonic transition energy) as a
function of (a) the substrate dielectric constant with vacuum above ($\eps_3 =
1$), and (b) the encapsulating dielectric constant ($\eps_2=\eps_3$).  The bulk
band gap, which is a fixed parameter in the theory, is indicated by a dotted
grey line.  Filled circles at $\eps_2 = 1$ indicate the ab initio $G_0W_0$
result (black circle) and the Bethe-Salpeter equation results (red and blue
circles) for an isolated monolayer, from Ref.~\onlinecite{Qiu2016}.
}
\label{fig:bulk_diff}
\end{figure}

First, we consider the experimentally-relevant situation of a monolayer
on a substrate with dielectric constant $\eps_2$ and vacuum above
($\eps_3 = 1$). In Fig.~\ref{fig:bulk_diff}(a), we show the band gap
calculated using the tight-binding
COHSEX approximation, as a function of the substrate dielectric constant.
The purely electrostatic approximation in Eq.~(\ref{eq:electro}) is not shown,
but gives nearly identical
results, predicting band gaps that are slightly larger (about 0.05~eV), which
can be understood based on arguments presented above.  
Remarkably, the simple theory presented here -- parameterized
only on bulk data and an estimate of the monolayer width -- predicts an isolated
monolayer ($\eps_2=1$) band gap of 2.62~eV (a 0.64~eV increase from bulk); this
compares very favorably to a recent, carefully-converged ab initio calculation using
the many-body $G_0W_0$ approximation,
which predicts 2.67~eV~\cite{Qiu2016}.  This huge increase in the quasiparticle
band gap reflects the strong role played by reduced dielectric screening
in atomically-thin materials.

At larger values of $\eps_2$, the increased
screening ability of the substrate yields a rapid decrease in the band gap,
demonstrating the strong sensitivity of atomically-thin materials to their local
environment.  Even a modest substrate like silica, with a dielectric
constant of $\eps_2 \approx 4$, is predicted to have a band gap of 2.35~eV, which is
0.27~eV smaller than an ideal, suspended monolayer.  On graphite, with $\eps_2\approx 10$,
the band gap is reduced by 0.45~eV. 
Similar results have been obtained with an approximate treatment of substrate
screening in otherwise ab initio $G_0W_0$ calculations~\cite{Ugeda2014,Ryou2016}.
These findings underscore
the care required when comparing experimental measurements on substrates to
ab initio calculations of isolated atomically-thin materials.
In reverse, the simple formula given in Eq.~(\ref{eq:electro}) can be used
to infer the ideal, suspended band gap based on measurements performed
on substrates.

In Fig.~\ref{fig:bulk_diff}(a), we also show the optical gap for the 1s and 2s
exciton states, obtained by summing the quasiparticle band gap and the
exciton binding energies of each state, as a function of the substrate dielectric
constant.  For the isolated monolayer,
we predict optical gaps of 2.03~eV and 2.35~eV (positive binding energies of
0.59~eV and 0.27~eV) for the 1s and 2s states, respectively.
Again, these compare well with converged ab initio calculations using the
Bethe-Salpeter equation, which predict optical gaps of 2.04~eV and 2.32~eV
(binding energies of 0.63~eV and 0.35~eV)~\cite{Qiu2016}.

As the dielectric constant of the substrate increases,
the exciton binding energies are reduced due to increased environmental screening.
Remarkably, the competing effects in the band gap and 1s binding energy
almost exactly cancel.  Up to a substrate dielectric constant of $\eps_2=20$,
the 1s optical transition energy only changes by 0.1~eV.  In the aforementioned
examples of silica and graphite substrates, the exciton binding energy is reduced by
0.24~eV and 0.49~eV, respectively. Not only is the optical
transition energy roughly constant, but the cancellation is almost perfect such
that the monolayer transition energy is nearly identical to the bulk transition
energy (the bulk band gap and optical gap roughly coincide, because the exciton
binding energy is only about 0.04~eV~\cite{Evans1965}).

In addition to the well-known observation that the optical gap of bulk TMDCs is
almost identical to that of monolayers, the effects predicted by the theory are 
in good agreement with a number of other more detailed experimental 
findings, such as the insensitivity of the optical gap in TMDCs when
comparing suspended samples and samples on fused silica substrates~\cite{Li2014}.
Identical effects in the band gap, optical gap, and exciton binding energy
have been observed in a joint experimental-computational study of MoSe$_2$
on bilayer graphene and graphite: the latter exhibits a 0.24~eV reduction
in the band gap and a concomitant 0.28~eV reduction in the exciton binding 
energy, leading to a minimal change in the optical gap~\cite{Ugeda2014}.

The above analysis can be repeated for more general dielectric environments;
the results of uniform encapsulation ($\eps_2=\eps_3$) are shown in
Fig.~\ref{fig:bulk_diff}(b).  While the qualitative behavior is the same,
the effects are naturally stronger due to the simultaneous screening from
above and below the monolayer.

Finally, we mention that although we have focused on the band gap, our theory
separately predicts changes to the ionization potential and electron affinity.
The environmental renormalization of these quantities may be of interest for
photochemistry, catalysis, or device engineering.

\textit{Conclusions.} In summary, we have presented a simple, but powerful
theory of environmentally-sensitive electronic and optical transition energies
in atomically-thin materials.  While the theory shows that the quasiparticle
band gap and the exciton binding energy are individually very sensitive to
their local dielectric environment, the sum of the two (the lowest-energy
optical transition) is almost completely insensitive.  In some sense, this is
an unfortunate state of affairs for the use of atomically-thin materials as
environmental or chemical sensors, because optical transitions are the simplest
to measure (by absorption or photoluminescence); by contrast, measuring the
band gap by photoemission or electron tunneling experiments is much more
difficult.  Nonetheless, the theory presented here enables rapid and
quantitative exploration of accessible energetic changes through dielectric
engineering.

In light of our results, we propose that the higher-lying excitonic resonances
are promising optical reporters of the local environment.  Even
the 2s resonance -- which can typically be resolved in experiments -- is
predicted to redshift by 0.1~eV when a suspended sample is placed on a silica
substrate.  Indeed, the 1s-2s separation was used recently as an experimental
probe of environmental effects~\cite{Raja2017}.

Going forward, this approach can be used to study other
environmentally-sensitive, atomically-thin materials such as black
phosphorous~\cite{Qiu2017}.  These techniques can also be applied to more
heterogeneous dielectric environments, as might be experimentally realized
through patterning~\cite{Raja2017}, molecular
coverage~\cite{Peimyoo2014,Feierabend2017}, or functional layered
heterostructures~\cite{Novoselov2016,Lin2016,Wilson2017}. In many cases,
explicit electronic hybridization and charge transfer should be accounted for
in the theory.  Work along these lines is currently in progress.

\vspace{1em}

This work was primarily supported by the University of Chicago Materials Research
Science and Engineering Center, which is funded by the National Science Foundation
under Award Number DMR-1420709.

\end{document}